\documentclass[aps,prx,groupedaddress,twocolumn,showpacs,longbibliography,10pt,nofootinbib]{revtex4-2}
\DeclareFontShape{OT1}{cmr}{mx}{n}{<->cmr10}{}
\fontseries{mx}\selectfont
\usepackage{color}
\usepackage{bm} 
\usepackage{units}
\usepackage{bbm}
\usepackage[caption=false]{subfig}
\usepackage{graphicx}
\usepackage{dsfont}
\usepackage{amsmath}
\usepackage{amssymb}
\usepackage{array}
\usepackage{epstopdf}
\usepackage{enumerate}
\usepackage{youngtab}
\usepackage{tensor}
\usepackage{braket}
\usepackage[normalem]{ulem}
\usepackage{slashed}
\usepackage[aligntableaux=center]{ytableau}
\usepackage[utf8]{inputenc}
\usepackage[
      colorlinks=true,
      linkcolor=blue,
     urlcolor=blue,
      filecolor=black,
      citecolor=red,
      ]{hyperref}
\usepackage{cleveref}
\usepackage{braket}
\usepackage{mathtools}
\usepackage{rotating}
\usepackage{tabularx}
\newcolumntype{Y}{>{\centering\arraybackslash}X}
\newcolumntype{C}[1]{>{\centering\arraybackslash}p{#1}}
\usepackage{todonotes}
\usepackage{xcolor}
\definecolor{LightCyan}{rgb}{0.7,1,1}
\definecolor{Gray}{gray}{0.9}
\usepackage{xspace}
\usepackage{slashed}

\begin{document}

\title{Precision study of the massive Schwinger model near quantum criticality}

\author{Erick Arguello Cruz, Grigory Tarnopolsky and Yuan Xin}
\affiliation{Department of Physics, Carnegie Mellon University, Pittsburgh, PA 15213, USA}

\begin{abstract}
We perform a numerical analysis of the massive Schwinger model in the presence of a background electric field. Using the Density Matrix Renormalization Group (DMRG) approach, we efficiently compute the spectrum of the Schwinger model on a staggered lattice with up to 3000 qubits. As a result, we achieve a precise computation of the critical mass of the massive Schwinger model to five digits using four different “criticality criteria,” observing perfect agreement among them. Additionally, we discuss the effect of a four-fermion operator deformation of the Schwinger model and compute the critical mass for various values of the deformation parameter.

\end{abstract}

\maketitle
\nopagebreak

\section{Introduction and Summary}
The Schwinger model \cite{Schwinger:1962tp, Lowenstein:1971fc, PhysRevD.10.732}
serves as an ideal $1+1$ dimensional toy model laboratory for the theoretical analysis and numerical testing of numerous Quantum Field Theory (QFT) phenomena, including the confinement of charged particles, chiral symmetry breaking, and second-order phase transitions.
The Lagrangian density of this model, in the presence of a background electric field parametrized by the $\theta$ angle, is given by \cite{Coleman:1975pw, Coleman:1976uz}
\begin{align}
    \mathcal{L} = -\frac{1}{4}F_{\mu\nu}F^{\mu\nu} -\frac{e\theta}{4\pi}\epsilon^{\mu\nu}F_{\mu\nu}+\bar{\psi}( i\slashed{\partial}- e\slashed{A} -m)\psi \,,
    \label{Schw_lagrange}
\end{align}
where $\psi$ is a two-component fermionic field, $\slashed{A} = \gamma^{\mu}A_{\mu}$,  where $(\gamma^{0},\gamma^{1})=(Z, iY)$ and $X,Y,Z$ are the Pauli matrices,  $F_{\mu\nu}=\partial_{\mu}A_{\nu}-\partial_{\nu}A_{\mu}$ is the  field strength,  $\epsilon^{01}=1$ and $e$ is the electric charge.
Using the timelike axial gauge  $A_{0}=0$ we find $F^{10}=-\partial_{t}A^{1}=E$, where $E$ is the electric field. We then obtain the Hamiltonian density of this model in the form:
\begin{align}
\mathcal{H} =  \frac{1}{2}\Big(E + \frac{e\theta}{2\pi} \Big)^2 +\bar{\psi}( m - i\gamma^1 (\partial_{1} +  i e A_{1}))\psi
\end{align}
and $[E(x), A_{1}(y)]=-i\delta(x-y)$. Using bosonization, the Hamiltonian density of the Schwinger model can be expressed as \cite{Coleman:1976uz}:
\begin{align}
\mathcal{H} = & N_{M_{\textrm{S}}}\Big[\frac{1}{2} \Pi^2 +\frac{1}{2}(\partial_{x}\phi )^2 +\frac{1}{2} M_{\textrm{S}}^2\phi^2 \notag\\
&\qquad\qquad -c m M_{S} \cos(2\sqrt{\pi}\phi -\theta)\Big]\,, 
\end{align}
where $c=e^{\gamma}/(2\pi) \approx 0.2835 $,  $M_{\textrm{S}}=e/\sqrt{\pi}$ is the Schwinger boson mass and $N_{M_{\textrm{S}}}$ denotes normal ordering with respect to mass $M_{\textrm{S}}$. For a background electric field with $\theta = \pi$, the effective potential of the bosonic field $\phi$ develops two minima when the mass $m$ is sufficiently large \cite{Coleman:1976uz}, indicating a second-order phase transition and the spontaneous breaking of the $Z_{2}$ symmetry $\phi \to -\phi$. As a result, there exists a critical mass $m_{c}/e$ at which this second-order phase transition occurs. It is worth studying this phase transition carefully, as a similar transition occurs in 4D QCD with one flavor \cite{Creutz:2006ts}.

In this paper, we aim to precisely determine the critical mass $m_{c}/e$ of the Schwinger model in background electric field corresponding to $\theta= \pi$ using numerical computations. One of the most accurate results for $m_{c}/e$ was obtained numerically in \cite{Byrnes:2002gj, byrnes2003density}. 
Other studies have also computed the critical mass using various methods, see e.g. \cite{PhysRevD.95.094509, Ohata:2023gru, Dempsey:2023gib}.  Our improved estimate is
\begin{align}
    m_{c}/e=0.333561(4).
\end{align}
Thus, our work decisively excludes the possibility that $m_c/e$ is exactly $1/3$.

The massive Schwinger model is not exactly solvable. However, despite its non-local Coulomb interaction, it can be efficiently analyzed numerically. Even the method of exact diagonalization can yield accurate results for small lattices \cite{Dempsey:2022nys}, provided a lattice version with the shifted mass is used, as discussed below. Additionally, many Monte Carlo computations have been performed for the Schwinger model, see e.g. \cite{Ranft:1982bi, Schiller:1983sj, Ohata:2023sqc, Ohata:2023gru}.

On the other hand, for $1+1$ dimensional models, the Density Matrix Renormalization Group (DMRG) method \cite{PhysRevLett.69.2863, PhysRevB.48.10345} can be applied to lattices of extremely large sizes or even infinite size \cite{PhysRevLett.75.3537, Vidal:2006ofj, McCulloch:2008aun, Zou:2017zce, PhysRevLett.118.220402}, typically providing excellent accuracy. The first application of DMRG to study the massive Schwinger model was presented in \cite{Byrnes:2002gj}. Subsequently, there have been numerous numerical investigations of the Schwinger model using DMRG (MPS) and more general Tensor Network methods, see e.g. \cite{Banuls:2013jaa, Banuls:2013zva, Shimizu:2014fsa, PhysRevLett.113.091601, Banuls:2016lkq, Buyens:2015tea, Buyens:2016ecr, Shimizu:2017onf, PhysRevD.101.054507, Honda:2021aum, Dempsey:2023gib, Itou:2024psm}.

In this article we perform the DMRG study of the Schwinger model using both the publicly available iTensor library\cite{itensor, itensor-r0.3} and our DMRG code implemented in MATLAB. This approach allows us to cross-check the numerical results and more accurately evaluate their precision.

The paper is organized as follows: In Section II, we define the Schwinger model on a staggered lattice, discuss various boundary conditions, and examine its deformation by the quartic operator. In Section III, we explain methods for locating the critical point of the second-order phase transition using the Hamiltonian approach. In Section IV, we analyze the structure of the energy levels of the Schwinger model Hamiltonian under different boundary conditions. In Section V, we present our numerical results for the critical mass of the Schwinger model with Open Boundary Conditions (OBC). In Section VI, we provide our numerical results for the critical mass of the Schwinger model with Periodic Boundary Conditions (PBC). In the appendix, we discuss the conformal perturbation theory approach used to analyze results in the case of PBC, along with details of the numerical implementation of the computations.

\section{Schwinger model on a lattice}
\label{Sec2}
\begin{figure}[h!]
\includegraphics[width=0.40\textwidth]{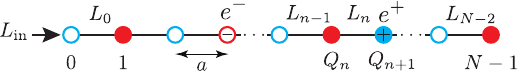}
\caption{Schwinger model on a lattice with open ends (OBC).}
\label{Schwinger_chain}
\end{figure}

There are various methods to discretize the Schwinger model, and one of the most common methods that avoids fermion doubling is the staggered lattice discretization introduced in \cite{Kogut:1974ag}. The discretized Hamiltonian of the massive Schwinger model on an open chain of $N$ sites is given by \cite{PhysRevD.13.1043}:
\begin{align}
H_{\textrm{S}} =& \frac{e^{2}a}{2} \sum_{n=0}^{N-2}\Big(L_{n}+\frac{\theta}{2\pi}\Big)^{2} + m_{\textrm{lat}} \sum_{n=0}^{N-1}(-1)^{n}c_{n}^{\dag}c_{n} \notag\\
&-\frac{i}{2a}\sum_{n=0}^{N-2}(c_{n}^{\dag}e^{i \phi_{n}} c_{n+1} - c_{n+1}^{\dag} e^{-i\phi_{n}}c_{n})\,, \label{SchLatt}
\end{align}
where $a$ is the lattice spacing, $L_{n}=-i \partial /\partial \phi_{n}$ and $\phi_{n}$ are the gauge field link variables, $c_{n}, c_{n}^{\dag}$ are  fermion annihilation and creationg operators at  site $n$ and we assume that $N$ is even. The discretized fermion creation and annihilation operators satisfy $\{c_{n},c_{m}^\dag\} = \delta_{nm}$. It was noted in \cite{Dempsey:2022nys} that in order to endow the lattice Hamiltonian with a certain symmetry for $m=0$ one has to introduce the lattice mass $m_{\textrm{lat}}=m -e^{2}a/8$. In the continuum limit ($a\to 0$)  $m_{\textrm{lat}}$ coincides with $m$. The Gauss's law can be written in the form \cite{Hamer:1997dx}
\begin{align}
L_{n}-L_{n-1} = Q_{n}, \quad Q_{n} = c_{n}^{\dag}c_{n} - \delta_{n,\textrm{odd}}
\end{align}
and it completely fixes the gauge field degrees of freedom in terms of the fermionic charges:
\begin{align}
L_{n} = \sum_{k=0}^{n}Q_{k} + L_{\textrm{in}}\,,
\end{align}
where $L_{\textrm{in}}$ is the incoming electric field from the left of the chain and we set it to zero $L_{\textrm{in}}=0$. Thus the gauge field states are completely determined by the fermionic ones and therefore the link variables $e^{i\phi_{n}}$ can be simply omitted in the Hamiltonian (\ref{SchLatt}). The Schwinger model lattice is schematically depicted in the Figure \ref{Schwinger_chain}.

The Jordan-Wigner transformation 
\begin{align}
c_{n} = \prod_{k=0}^{n-1}(iZ_{k}) \otimes  \frac{1}{2}(X_{n}-iY_{n})\,,
\end{align}
allows to write the Schwinger model Hamiltonian in spin variables as:
\begin{align}
\frac{2}{e^{2}a}H_{\textrm{S}} = H_{C} + \mu H_{\textrm{M}} + x H_{\textrm{XX}} \,, \label{LatticeSchM}
\end{align}
where $x= 1/(e^2a^2)$, $\mu=2\sqrt{x} m_{\textrm{lat}}/e$ and the Coulomb energy term $H_{C}$, the mass term $H_{\textrm{M}}$ and the  XX term $H_{\textrm{XX}}$  read: 
\begin{align}
&H_{\textrm{C}} =  \sum_{n=0}^{N-2}\Big( \frac{1}{2}\sum_{k=0}^{n}(Z_{k}+ (-)^{k}) +\frac{\theta}{2\pi} \Big)^2\,, \label{HCOBC}\\
&H_{\textrm{M}} = \frac{1}{2}\sum_{n=0}^{N-1}(-1)^{n}Z_{n}\,, \label{HMOBC} \\
&H_{\textrm{XX}} = \frac{1}{2} \sum_{n=0}^{N-2}(X_{n}X_{n+1}+Y_{n}Y_{n+1})\,. \label{HXXOBC}
\end{align}
The $U(1)$ charge reads $Q =\frac{1}{2}\sum_{n=0}^{N-1}Z_{n}$ and it commutes with the  Hamiltonian $[H_{\textrm{S}},Q]=0$. The Hamiltonian $H_{\textrm{S}}$ acts in the $2^{N}$ dimensional Hilbert space spanned by a basis of $N$-qubit states $\mathcal{H}_{\textrm{S}}=\{|\sigma_{0}\sigma_{2}\dots \sigma_{N-1}\rangle,\, \sigma_{i}=\uparrow, \downarrow\}$.

The Schwinger model defined in (\ref{Schw_lagrange}) can be deformed by adding a marginal four-fermion operator
$O = \lambda (\bar{\psi} \psi)^2$
to the Lagrangian  \cite{Komargodski:2020mxz, Cherman:2022ecu}\footnote{This operator is not generated by staggered lattice discretization but rather has to be added by hand. We also notice that for a single two component spinor in two dimensions 
we have an equality $(\bar{\psi} \psi)^2  = (\bar{\psi}\gamma^{5} \psi)^2 = -1/2  (\bar{\psi}\gamma^{\mu} \psi)^2 $ due to the Fierz identity \cite{Bondi:1989nq}.}. Such a model was dubbed Schwinger-Thirring (ST) model  and this deformation changes the lattice Hamiltonian term $H_{\textrm{XX}}$ to   $H_{\textrm{XXZ}}$  \cite{PhysRevD.16.3031, Banuls:2018ckt}:
\begin{align}
&H_{\textrm{XXZ}} = \frac{1}{2} \sum_{n=0}^{N-2}(X_{n}X_{n+1}+Y_{n}Y_{n+1} -\frac{\lambda}{2}Z_{n}Z_{n+1})\,. \label{HXXZ}
\end{align}
It is well known that DMRG computations for systems with open boundary conditions (OBC) are significantly faster than those for periodic boundary conditions (PBC) \cite{PhysRevLett.69.2863, PhysRevB.48.10345}. On the other hand, systems with PBC exhibit translational symmetry and are free from boundary effects. In many systems with PBC, away from a critical point, convergence to the thermodynamic limit occurs exponentially fast with the number of lattice sites $N$ \cite{hamer1982massive}.

\begin{figure}[h!]
\includegraphics[width=0.15\textwidth]{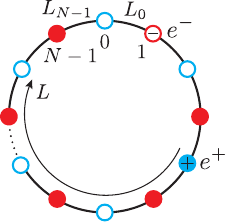}
\caption{Schwinger model on a lattice with closed ends (PBC).}
\label{Schwinger_circle}
\end{figure}

In the case of PBC for the Schwinger model all the fermionic quantum states must have zero $U(1)$ charge $\sum_{k} Q_{k} =0$. 
Moreover, one has to take into account an additional gauge field  holonomy mode $L$,  which can take any integer value and is not fixed by the charges $Q_{k}$ on the lattice sites, see FIG. \ref{Schwinger_circle}.
In this case the Coulomb Hamiltonian $H_{\textrm{C}}$ in (\ref{HCOBC}) is modifed to 
\begin{align}
&H_{\textrm{C}} =  \sum_{n=0}^{N-1}\Big( \frac{1}{2}\sum_{k=0}^{n}(Z_{k}+ (-)^{k}) +\frac{\theta}{2\pi} + L \Big)^2\,, \label{HCPBC}
\end{align}
and the XX Hamiltonian $H_{\textrm{XX}}$ in (\ref{HXXOBC}) has an extra boundary term \cite{PhysRevD.99.094501}:
\begin{align}
&H_{\textrm{XX bdry}} =  \sigma_{N-1}^{+}\sigma_{0}^{-}L_{+} + \sigma_{N-1}^{-}\sigma_{0}^{+}L_{-}\,,
\end{align}
where $\sigma^{\pm} = (X \pm i Y)/2$ and $[L, L_{\pm}] = \pm L_{\pm}$. The Hilbert space is  $\mathcal{H}_{\textrm{S}}=\{ |\sigma_{0}\sigma_{2}\dots \sigma_{N-1}\rangle \otimes |\ell\rangle,\, \sigma_{i}=\uparrow, \downarrow;\, \ell \in \mathbbm{Z}\}$, where the states $|\ell\rangle$ satisfy $L|\ell\rangle = \ell |\ell\rangle$ and $L_{\pm}|\ell\rangle = |\ell \pm 1\rangle$.
For numerical computations we truncate the holonomy space and assume that $\ell$ only takes a finite number of values $\ell = -\ell_{\textrm{max}},\dots, \ell_{\textrm{max}}$.

\section{Locating the critical point}
The ferromagnetic critical point of the transverse field Ising model  with the Hamiltonian 
\begin{align}
&H_{\textrm{Ising}} = -\sum_{n=0}^{N-2}Z_{n}Z_{n+1} - g \sum_{n=0}^{N-1}X_{n} \label{IsingH}
\end{align}
is located at $g_{c} = 1$ and this can be inferred from the Kramers-Wannier duality, which relates the Ising Hamiltonian with the field strength $g$ to the one with $1/g$ \cite{PhysRevD.17.2637}.  It has been suggested that the ferroelectric critical point of the massive Schwinger model at $\theta =\pi$ belongs to the 2d Ising model universality class \cite{Byrnes:2002gj, Dempsey:2022nys, Ohata:2023gru}. 
But in  the  Schwinger model there is no apparent Kramers-Wannier duality and therefore in general we expect the critical point $m_{c}/e$ to be some irrational number. Moreover the lattice Schwinger model (\ref{LatticeSchM}) has a critical point for any value of the coupling $x$. Therefore $m_{c}/e=m_{c}(x)/e$ is itself a function of $x$ and to compute the critical mass in the continuum  we have to take the limit $x\to \infty$ at the end:
\begin{align}
    m_{c}/e=\lim_{x\to\infty}m_{c}(x)/e\,.
\end{align}

In order to locate the critical 
point numerically for given $x$ we use the finite size scaling hypothesis \cite{PhysRevLett.28.1516}, which implies that at the critical point the energy gaps $E_{n0}\equiv E_{n}-E_{0}$ scale with the size of the system $N$  as  \cite{Hamer:1981qt}:
\begin{align}
E_{n0} (m_{c}/e, x, N)\propto \Delta_{n}/N. \label{FFSgaps}
\end{align}
Moreover, the coefficients $\Delta_{n}$ are the scaling dimensions of the operators at the critical point \cite{Cardy:1984epx}. Using this, we define a "pseudo-critical" coupling $m_{*}/e$ for a lattice of size $N$  as a point at which the following equality holds \cite{Hamer:1981qt, Byrnes:2002gj}:
\begin{align}
C_{1}: \quad \frac{(N+2) E_{10}( m_{*}/e, x, N+2)}{N E_{10}(m_{*}/e, x, N)} = 1\,, \label{CritCr1}
\end{align}
where the gaps $E_{10}$ in the numerator and denominator are computed for the lattices of sizes $N+2$ and $N$ respectively\footnote{We grow the system by two sites since on the staggered lattice they represent one discrete space point}. The equation (\ref{CritCr1}) defines the "pseudo-critical" coupling $m_{*}(x, N)/e$ as a function of the lattice size $N$. In the limit of infinite $N$ such defined coupling $m_{*}(x, N)/e$ will approach the critical value of the coupling in thermodynamic limit: 
\begin{align}
m_{c}(x)/e = \lim_{N \to \infty} m_{*}(x,N)/e  \,.
\end{align}
Therefore in order to numerically find $m_{c}(x)/e$ we need to compute $m_{*}(x,N)/e$ for a sequence of values $N$ using  (\ref{CritCr1}) and then compute the approximation of $m_{*}(x,N)/e$ for $N \to \infty$.

We refer to the relation (\ref{CritCr1}) as the "criticality criterion". Using the prediction of the finite size scaling (\ref{FFSgaps}) it is possible to construct infinitely many different "criticality criteria". However, the most practical criterion is the one for which the function $m_{*}(N)$ converges to $m_{c}$ most rapidly. 
In this article we use four different "criticality criteria". We call them $C_{i}$, with $i=1,2,3,4$ and the first criterion $C_{1}$ is defined in (\ref{CritCr1}). The criterion $C_{1}$ doesn't rely on any information about the universality class of the critical point; but for the next criteria, we use that it is 2D Ising Conformal Field Theory (CFT).
The second criterion is defined as
\begin{align}
C_{2}: \quad \frac{E_{20}(m_{*}/e, x,N)}{E_{10}(m_{*}/e, x,N)} =  \frac{\Delta_{2}}{\Delta_{1}}\,,
\label{CritCr2}
\end{align}
where $\Delta_{1}$ and $\Delta_{2}$ are the exact values of the scaling dimensions in the 2D Ising CFT \cite{Belavin:1984vu} and they do not depend on $x$. The scaling dimensions $\Delta_{1}$ and $\Delta_{2}$ depend on the boundary conditions of the quantum Hamiltonian, but in all the cases they are known exactly for the 2D Ising CFT \cite{Cardy:1986gw}. For the Hamitlonian (\ref{LatticeSchM}) with OBC we have $\Delta_{2}/\Delta_{1}=3/2$ and for the Hamltonian with PBC we have $\Delta_{2}/\Delta_{1}=8$ (we discussed it in the sec. \ref{SecEnLevs} ). 

For the third and fourth criteria $C_{3}$ and $C_{4}$ we use the von Neumann entanglement entropy $S(N_{L})$ of the left subsystem with $N_{L}$ sites ($N_{L}+N_{R}=N$):
\begin{align}
S(N_{L}) = - \textrm{Tr}(\rho_{L} \ln \rho_{L})\,,
\end{align}
where $\rho_{L}$ is the left density matrix  
\begin{align}
\rho_{L}=\textrm{Tr}_{R}(|\psi_{0}\rangle \langle \psi_{0}|)\,,
\end{align}
and $|\psi_{0}\rangle$ is the ground state of the Hamiltonian.
In this case the expression of the entanglement entropy at the critical point for a lattice of $N$ sites and its left subsystem of $N_{L}$ sites   is \cite{Calabrese:2004eu, Calabrese_2009}:
\begin{align}
    S(N_{L})&=\frac{\kappa c}{6}\log{\left(\frac{2N}{\kappa \pi}\sin{\frac{\pi N_{L}}{ N}}\right)}+g(2-\kappa)+c_{1}'\,,
    \label{eq_entropy}
\end{align}
where $c$ is the central charge of the 2D CFT and  $\kappa=2$ corresponds to PBC and $\kappa=1$ for OBC, $g$ is the boundary entropy \cite{PhysRevLett.67.161} and $c_{1}'$ is a non-universal constant.  Away from the critical point $S = (\kappa c  /6)\ln (\xi/a)$, for $N_{L},N\gg \xi/a$, where $\xi$ is the correlation length and $a$ is the lattice spacing. 

Following \cite{Campostrini_2014} we  define a function:
\begin{align}
    \mathcal{S}(N_{L},N'_{L}; N)=\frac{6}{\kappa} \frac{S(N_{L})-S(N_{L}')}{\log{\left( \sin{\frac{\pi N_{L}}{N}}/\sin{\frac{\pi N_{L}'}{N}}\right)}}.
\end{align}
Then at the critical point
\begin{align}
    \lim_{N\to\infty}\mathcal{S}(N_{L},N_{L}';N)=c\,,
    \label{eq_limQ}
\end{align}
otherwise for $N \gg \xi/a$ we should find $\mathcal{S}(N_{L},N_{L}';N)=0$. 

In our case, we take $N_{L}=N/2$, $N_{L}'=N/2-4$, and we define $m_{*}/e$ as the point where:
\begin{align}
  C_{3}: \quad \mathcal{S}(N/2,N/2-4;N)= c\,, \label{eq_q}
\end{align}
and for the 2D Ising critical point $c=1/2$.
The last criticality criterion we use is 
\begin{align}
  C_{4}: \quad  \frac{\mathcal{S}(N/2+2, N/2-2; N+4)}{\mathcal{S}(N/2, N/2-4; N)}
    = 1\,. \label{eq_q}
\end{align}
We notice that the criticality criteria $C_{1}$ and $C_{4}$ do not rely on the knowledge of the critical point universality class.

\section{Energy levels of the Schwinger model at the critical point}
\label{SecEnLevs}
\begin{figure}[h!]
\includegraphics[width=\linewidth]{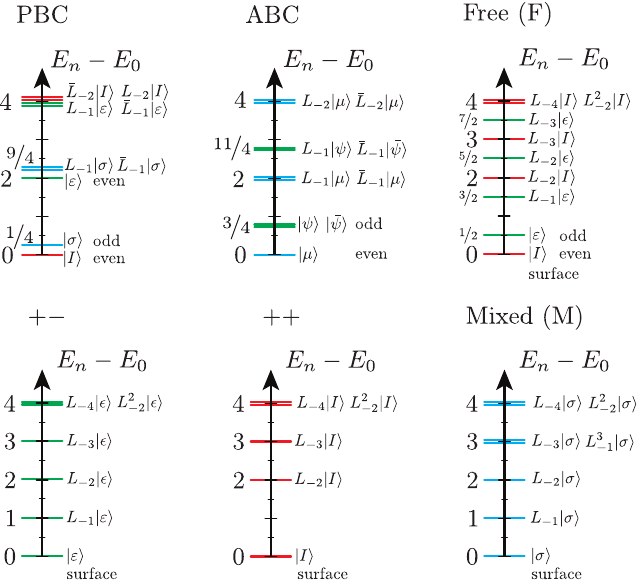}
\caption{Energy gaps for 2D Ising CFT (\ref{IsingH}) with different boundary conditions in units of $2\pi/N$.}
\label{IsingCFT_BC}
\end{figure}

As stated in (\ref{FFSgaps}), the energy gaps of a model at its critical point are connected to the scaling dimensions of certain operators. The  operators associated with these energy levels depend on the boundary conditions of the model. The first six different boundary conditions and their scaling dimensions (energy levels) for the 2D Ising CFT (\ref{IsingH}) were provided in \cite{Cardy:1986gw}. These are Periodic (P), Antiperiodic (A), Free (F), $(++)$, $(+-)$ and Mixed (M) boundary conditions. Their energy gaps and corresponding Virasoro states are depicted in FIG. \ref{IsingCFT_BC}. 
The last four boundary conditions correspond to the surface operators. It is known that the Schwinger model at $\theta =\pi$ at the critical point belongs to the 2D Ising universality class, therefore we expect that the energy gaps of the Hamiltonian in (\ref{LatticeSchM}) for any $x$ coincide with the ones in the FIG. \ref{IsingCFT_BC} up to some overall factor. It is more convenient to use dimensonless ratios of the energy gaps $E_{k0}/E_{10} = \Delta_{k}/\Delta_{1}$ since they don't depend on the overall factor. Numerically we identified the following relations between the Schwinger model Hamiltonian with different boundary conditions and the 2D Ising CFT boundary conditions in FIG. \ref{IsingCFT_BC}:
\begin{align}
&H_{\textrm{S}}^{\textrm{OBC}}(\theta =\pi, Q = 0) \quad\;\;\; \leftrightarrow \quad (++)  \label{ScPP}\\
&H_{\textrm{S}}^{\textrm{OBC}}(\theta =\pi, Q =\pm 1) \quad \leftrightarrow \quad (+-)  \label{ScPM}\\
&H_{\textrm{S}}^{\textrm{PBC}}(\theta =\pi) \qquad\quad\quad\quad \leftrightarrow \quad \textrm{PBC} 
\end{align}
where $Q=0$ and $Q=\pm 1$ are the  $U(1)$ charge sectors.
For $(++)$ BC the ratio of energy gaps is $\Delta_{2}/\Delta_{1}=3/2$ and we used it in $C_{2}$ criterion in (\ref{CritCr2}).  Since in the Schwinger model the order parameter is the electric field, the relations (\ref{ScPP}) and (\ref{ScPM}) are expected since for $Q=0$ case we have $L_{\textrm{in}}=L_{\textrm{out}}$ and for $Q=\pm 1$ we have $L_{\textrm{in}}\neq L_{\textrm{out}}$
at the boundary. 

It was noticed in \cite{Byrnes:2002gj} that if one averages the energy levels of the Hamiltonians $H_{\textrm{S}}^{\textrm{OBC}}(\theta =\pi, -Q)$ and 
$H_{\textrm{S}}^{\textrm{OBC}}(\theta =-\pi, Q)$: 
\begin{align}
\frac{E_{\textrm{S}}^{\textrm{OBC}}(\theta =\pi, -Q)+E_{\textrm{S}}^{\textrm{OBC}}(\theta =-\pi, Q)}{2}\; \leftrightarrow \; \textrm{FBC}  \label{SpecAv}
\end{align}
one obtains a spectrum (combining all the charge sectors $Q$), which coincides with the one of the Ising model with Free BC. But in this case it corresponds to the dual Ising model, since this spectrum has first two energy levels separated by exponentially small gap for $m/e < m_{c}/e$, rather than for $m/e > m_{c}/e$ as in the PBC case. 
We noticed that parity $\bm{P}$ and the spin-inversion symmetry $\bm{Z} $:
\begin{align}
\bm{P} Z_{n}\bm{P} = Z_{N-n-1}, \quad  \bm{Z} Z_{n} \bm{Z} = -Z_{n}
\end{align}
imply the following relation between the Hamiltonians: 
\begin{align}
(\bm{P} \bm{Z})H_{\textrm{S}}^{\textrm{OBC}}(\theta , Q) (\bm{P} \bm{Z}) =
H_{\textrm{S}}^{\textrm{OBC}}(\theta -2\pi Q , Q)
\end{align}
in the given $U(1)$ charge sectors. Since $\bm{P} \bm{Z}$ flips the $U(1)$ charge we obtain the following equality of spectra 
\begin{align}
E_{\textrm{S}}^{\textrm{OBC}}(\theta , -Q)  =
E_{\textrm{S}}^{\textrm{OBC}}(\theta -2\pi Q , Q)\,.
\end{align}
Particularly this implies that $E_{\textrm{S}}^{\textrm{OBC}}(\theta = \pi , Q=-1) = 
E_{\textrm{S}}^{\textrm{OBC}}(\theta = -\pi , Q= 1)$. We notice that at  $m_{\textrm{lat}}=0$ ($\mu=-1/4$) the Hamiltonian $H_{\textrm{S}}^{\textrm{OBC}}$ does not contain $\sum_n (-1)^n Z_{n}$ term \cite{PhysRevD.107.054506} and thus  $\bm{P}$ and $\bm{Z}$ operators independently lead to different relations between the Hamiltonians $H_{\textrm{S}}^{\textrm{OBC}}(\theta, Q)$: 
\begin{align}
\bm{P} H_{\textrm{S}}^{\textrm{OBC}}(\theta , Q) \bm{P} =&
H_{\textrm{S}}^{\textrm{OBC}}(-\theta-\pi -2\pi Q , Q) \notag\\
& + \frac{Q}{2}+\frac{\theta}{2\pi} + \frac{1}{4}\,, \\
\bm{Z} H_{\textrm{S}}^{\textrm{OBC}}(\theta , Q) \bm{Z} =&
H_{\textrm{S}}^{\textrm{OBC}}(-\theta-\pi , Q) \notag\\
& -\frac{Q}{2}+\frac{\theta}{2\pi} + \frac{1}{4}\,.
\end{align}
This is an OBC analog of the symmetry enhancement at $m_{\textrm{lat}}=0$, first noticed in \cite{Dempsey:2022nys}.

\begin{figure}[h!]
\includegraphics[width=0.35\textwidth]{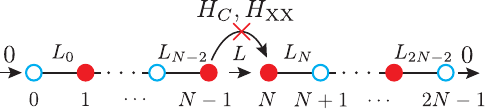}
\caption{An example of a Schwinger lattice with the spectrum at the critical point corresponding to the 2D Ising CFT with Free BC.  There are no electric energy $H_{C}$ and $H_{\textrm{XX}}$ terms between $N-1$ and $N$ sites, but the electric field outgoing from the first part of the chain enters its second part.}
\label{Schwinger_chain_average}
\end{figure}

The averaging procedure in (\ref{SpecAv}) is justified due to the asymmetry between positron and electron sites of the staggered lattice discretization (see also \cite{Honda:2022edn}), nevertheless it makes sense to identify a single Hamiltonian for the Schwinger model, such that its spectrum at the critical point coincides to the one of the 2D Ising CFT with Free BC. One such Hamiltonian is depicted in FIG. \ref{Schwinger_chain_average}. 
We assume that there are no electric energy $H_{C}$ and $H_{\textrm{XX}}$ terms between $N-1$ and $N$ sites, but the electric field outgoing from the first part of the chain enters its second part. The incoming electric field is zero: $L_{\textrm{in}}=0$ and the total $U(1)$ charge of the whole chain is zero as well. Such a Hamiltonian at $\theta=\pi$ has the energy spectrum which coincides with the one in (\ref{SpecAv}). It is convenient to use the spectrum of this Hamiltonian for the criterion $C_{1}$ defined in (\ref{CritCr1}).

\section{Numerical results for OBC}
We study the lattice Hamiltonian of the massive Schwinger model using the Density Matrix Renormalization Group (DMRG). We compute the energy levels of the Hamiltonian \eqref{LatticeSchM} at $\theta = \pi$ for $N\in[200,1000]$ in steps of $100$ and $x\in [10,100]$ in steps of 10\footnote{This is for  the criticality criteria $C_{1},C_{2}$ and $C_{4}$. For $C_{3}$ convergence is slower for smaller values of $N$, therefore we used $N\in[800,1000]$ in steps of 4. Additionally, to ensure this method converges to the correct value, we also included the $N=3000$ point, for all the values of $x$.}. 

\begin{figure}[htbp]
    \centering
    \includegraphics[width=\linewidth]{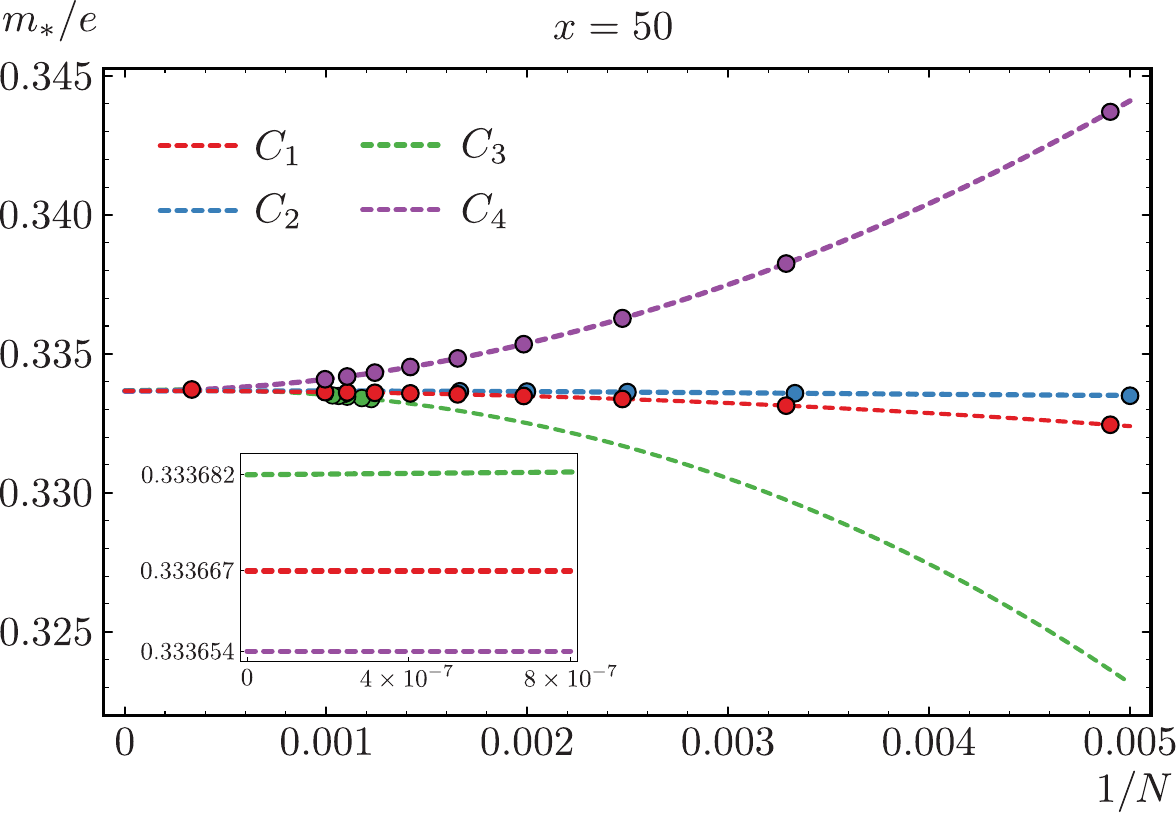} 
    \caption{Critical masses as a function of $1/N$ for $x=50$. The colored points are the values of $m_{*}(x, N)/e$ using the criticality criteria $C_{1},\dots, C_{4}$ and the dashed lines are the $1/N$ polynomial fits. Inset: Extrapolation values of the fittings when $N\to\infty$. We see that all the four methods converge to the same point within $10^{-5}$ error.}
    \label{fig:Summaryx50}
\end{figure}

To find the "pseudo-critical" point $m_{*}(x,N)/e$ for a fixed $N$ and $x$, we use a modified Newton's algorithm with a precision controlled by reaching the given criticality criteria $C_{1}, \dots, C_{4}$ up to $10^{-8}$ accuracy. For instance, for the $C_{1}$ case, we recursively vary $m(x, N)/e$ until the condition (\ref{CritCr1}) is fulfilled with $10^{-8}$ accuracy. The results for the four criticality criteria, for $x=50$ at different values of $N$ are shown in FIG. \ref{fig:Summaryx50} 
and their respective polynomial fittings in $1/N$ up to order 4 are\footnote{For the $C_{3}$ case, we only consider $1/N$ up to cubic order.} 
\begin{align*}
    C_{1}:&\quad m_{*}/e= 0.3336669 - \frac{41.70}{N^2} - \frac{3100}{N^3} + \frac{255730}{N^4}\,, \notag\\
    C_{2}:&\quad m_{*}/e= 0.3336670 - \frac{0.75}{N^2} - \frac{3985}{N^3} + \frac{566213}{N^4}\,, \notag\\
    C_{3}:&\quad m_{*}/e= 0.3336824 + \frac{0.2247}{N} - \frac{366.3}{N^2} - \frac{20025}{N^3}\,, \notag\\
    C_{4}:&\quad m_{*}/e= 0.3336540 + \frac{437.72}{N^2} - \frac{4134}{N^3} + \frac{38547}{N^4}\,. 
\end{align*}
We can roughly estimate an error of the results by analyzing the value of the last $1/N^{4}$ term, provided each subleading term is smaller than the previous ones. We see that $1/N$ convergence of $C_{1}$ and $C_{2}$ is better than  convergence of $C_{3}$ and $C_{4}$.

\begin{figure}[htbp]
    \centering
    \includegraphics[width=\linewidth]{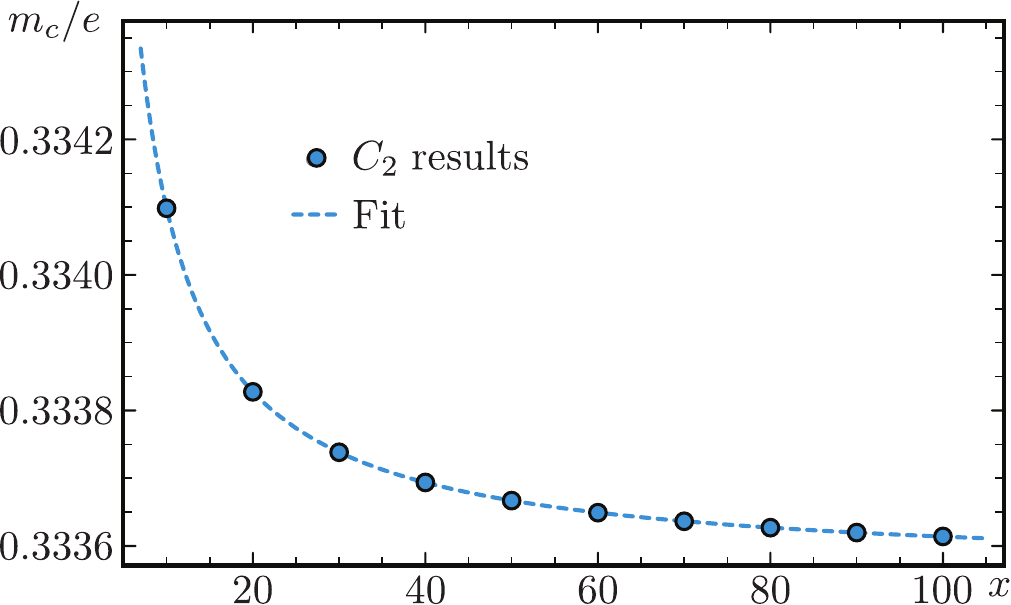} 
    \caption{Plot of the critical mass $m_{c}/e$ as a function of $x$, obtained using the $C_{2}$ criticality criterion. The $x=\infty$ value is $m_{c} /e=0.3335608$.}
    \label{fig:C2}
\end{figure}

Using the  $m_{c}(x)/e$ results obtained from $1/N$ fittings for different values of $x$, we obtain the extrapolated results as $x\to \infty$, by fitting them with a polynomial in $1/x$ up to cubic order. 
We summarize results for $m_{c}(x)/e$ for different values of $x$ and their  extrapolation to $x=\infty$ in the Table \ref{Table2}. We also included standard deviation of the four methods.
\begin{table}[htbp]
    \centering
\begin{tabular}{|c|c|c|c|c|c|}
\hline
& \multicolumn{4}{|c|}{$m_{c}(x)/e$} &  $\sigma_{m/e}$ \\
\hline
$x$ & $C_{1}$ &  $C_{2}$ & $C_{3}$ & $C_{4}$  &  \\
\hline
10 & $0.3340985$& $0.3340984$ & $0.3340950$  & $0.3340834$ & $6\cdot 10^{-6}$\\
\hline
20 & $0.3338274$& $0.3338274$ & $0.3338258$  & $0.3338111$ & $7\cdot 10^{-6}$\\
\hline
30 & $0.3337383$& $0.3337383$ & $0.3337449$  & $0.3337228$& $8\cdot 10^{-6}$\\
\hline
40 & $0.3336940$& $0.3336938$ & $0.3337042$  & $0.3336801$ & $9\cdot 10^{-6}$\\
\hline
50 & $0.3336669$& $0.3336670$ & $0.3336824$  & $0.3336540$ & $10^{-5}$ \\
\hline
60 & $0.3336490$& $0.3336492$ & $0.3336659$  & $0.3336379$ & $10^{-5}$\\
\hline
70 & $0.3336362$& $0.3336365$ & $0.3336391$  & $0.3336290$ & $3\cdot 10^{-6}$\\
\hline
80 & $0.3336266$& $0.3336270$ & $0.3336285$  & $0.3336246$ & $1\cdot 10^{-6}$\\
\hline
90 & $0.3336195$& $0.3336198$ & $0.3336345$  & $0.3336101$ & $9\cdot 10^{-6}$\\
\hline
100 & $0.3336126$& $0.3336141$ & $0.3336241$  & $0.3336083$ & $6 \cdot 10^{-6}$\\
\hline
$\infty$ & $0.3335585$ & $0.3335608$ & $0.3335593$  & $0.3335654$ & $4\cdot 10^{-6}$ \\
\hline
\end{tabular}
    \caption{The critical mass $m_{c}(x)/e$ for the four criticality criteria $C_{1},\dots, C_{4}$ and its standard deviation at different values of $x$, including extrapolation to $x=\infty$.}
    \label{Table2}
\end{table}
An example of the $x\to \infty$ extrapolation for the $C_{2}$ criterion is presented in FIG. \ref{fig:C2}.
We list $1/x$ fitting polynomials for $m_{c}(x)/e$:
\begin{align*}
    C_{1}:&\quad m_{c}/e =  0.3335585 + \frac{0.00548}{x} - \frac{0.0033}{x^2} + \frac{0.025}{x^3}\,, \notag\\
    C_{2}:&\quad m_{c}/e = 0.3335608 + \frac{0.00529}{x} + \frac{0.0008} {x^2} + \frac{0.0004}{ x^3}\,, \notag\\
    C_{3}:&\quad m_{c}/e = 0.3335593 + \frac{0.00665} {x} - \frac{0.0401}{x^2} + \frac{0.272}{x^3}\,, \notag\\
    C_{4}:&\quad m_{c}/e = 0.3335654 + \frac{0.00411}{x} + \frac{0.0217}{x^2} - \frac{0.111}{x^3}\,.
\end{align*}
We see that all of the results for the critical point coincide up to an error of order $10^{-6}$ with an average value of
\begin{align}
    m_{c}^{\textrm{av}}/e=0.333561(4).
\end{align}
We also observe that $1/x$ terms are quite small, which is due to the proper definition of the lattice mass $m_{\textrm{lat}}$ discussed in the Sec. \ref{Sec2}. Therefore the leading source of error is $1/N$ expansion.

\begin{figure}[htbp]
    \centering
    \includegraphics[width=\linewidth]{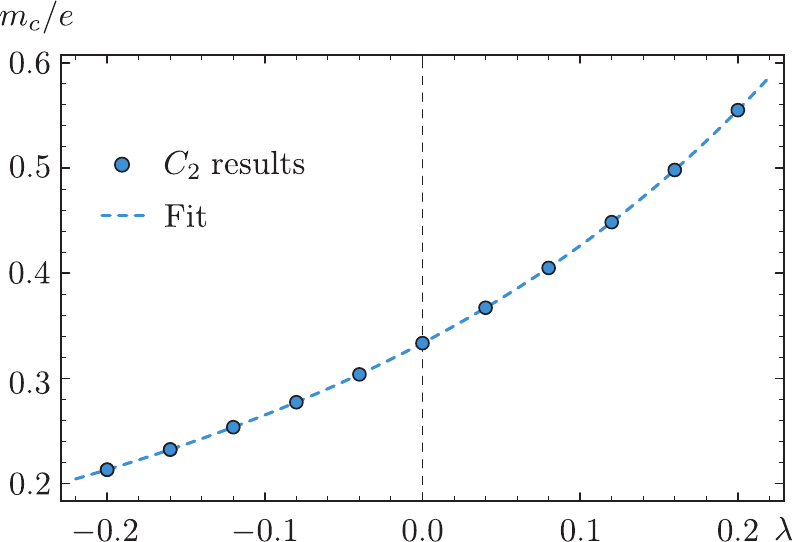} 
    \caption{Numerical results for the critical mass of the Schwinger-Thirring model as a function of $\lambda$. This results were obtained using the $C_{2}$ criticality criterion.}
    \label{fig:mvslambda}
\end{figure}

Finally we plot in FIG. \ref{fig:mvslambda} the value of the critical mass $m_{c}(\lambda)/e$ in the continuum model as a function of deformation parameter $\lambda$ introduced in (\ref{HXXZ}). For a range of $\lambda \in [-0.2,0.2]$, the polynomial fit reads
\begin{align}
    m_{c}(\lambda)/e&=0.333572 + 0.787592 \lambda + \notag\\ 
    & \quad  + 1.1911 \lambda^2 + 1.66328 \lambda^3 + 1.82977 \lambda^4,
\end{align}
where we have used the $C_{2}$ criterion with a Newton step error of $10^{-5}$.

\subsection{More on Conformal Spectrum of OBC and PBC}
At the critical point, the lattice spectrum is gapless, and for large but finite lattice size $N$, the $O(1/N)$ part of the spectrum agrees with the spectrum of CFT with the corresponding boundary condition, denoted as $X$, in space, up to a constant vacuum energy shift. The finite temperature partition function thus lives in a compactified Euclidean time with periodic boundary condition in time, we denote such a partition function as $Z_{PX}$ for spacial boundary condition $X$. The states in 2D CFT are representations of Virasoro algebra, and for 2D Ising CFT there are only 3 independent highest-weight representations, aka primary states, 
\begin{equation}
  \begin{array}{c|c}
  \hline
      \text{name} & \text{conformal weight } h  \\
      \hline
      \mathbbm{1} & 0  \\
      \hline
      \epsilon & \frac{1}{2}  \\ 
      \hline
      \sigma & \frac{1}{16}  \\ 
      \hline
  \end{array}
\end{equation}
and other states are completely determined by the Virasoro algebra, generating the Virasoro characters $\chi_{\mathcal{O}} = \sum_{i\in\mathcal{O}}q^{h_i}$
\begin{equation}
    \begin{aligned}
        \chi_{\mathbbm{1}} &= q^{-c/24}(1 + q^2 + q^3 + 2q^4 + \cdots) \\ 
        \chi_{\epsilon} &= q^{\frac{1}{2}-c/24}(1 + q + q^2 + q^3 + 2q^4 \cdots ) \\ 
        \chi_{\sigma} &= q^{\frac{1}{16}-c/24}(1 + q + q^2 + 2q^3 + 2q^4 + \cdots )
    \end{aligned}
\end{equation}
which are the building blocks of the partition function shown below. $c = 1/2$ is the central charge of Ising CFT and its only job is a constant shift of vacuum energy.
The partition functions further satisfy modular invariance, the invariance under swapping the space and Euclidean time followed by a rescaling. These conditions in turn pose strict constraints on the CFT spectrum and in the Ising CFT case completely determines the spectrum for all the boundary conditions in FIG. \ref{IsingCFT_BC} \cite{Cardy:1986gw}. We briefly summarize these results as follows.

The periodic ($P$) and anti-periodic ($A$) boundary condition have both spacial and time translation symmetry, and the states are representations of both left- and right- moving Virasoro algebras. The partition functions are 
\begin{equation}
    \begin{aligned}
        Z_{PP} &= \chi_{\mathbbm{1}}\bar\chi_{\mathbbm{1}} + \chi_{\sigma}\bar\chi_{\sigma} + \chi_{\epsilon}\bar\chi_{\epsilon} \\ 
        Z_{PA} &= \chi_{\sigma}\bar\chi_{\sigma}  + \chi_{\epsilon}\bar\chi_{\mathbbm{1}} +  \chi_{\mathbbm{1}}\bar\chi_{\epsilon}
    \end{aligned}
\end{equation}
with the identification $q = \bar q = e^{-\beta}$ and $E = h + \bar{h}$. In $P$ boundary condition the prediction is that there will be 3 towers of states built from primaries $\mathbbm{1}$, $\epsilon$ and $\sigma$. In $A$ boundary condition the prediction is that the parity-odd ground state $\mu$ has conformal weight $(\frac{1}{16}, \frac{1}{16})$, and there are two chiral fermion primaries $\psi\sim (\frac{1}{2},0)$ and $\bar\psi\sim (0,\frac{1}{2})$.  

The OBC family, consisting of free boundary condition ($F$), two fixed boundary conditions ($++$) and ($+-$), and the mixed boundary condition ($M$) do not have spacial translation symmetry, so they only have the diagonal Virasoro algebra. The partition functions are
\begin{equation}
    \begin{aligned}
        Z_{PF} &= \chi_{\mathbbm{1}} + \chi_{\epsilon} \\ 
        Z_{(++)F} &= \chi_{\mathbbm{1}}  \\ 
        Z_{(+-)F} &= \chi_{\epsilon} \\ 
        Z_{MF} &= \chi_{\sigma}
    \end{aligned}
\end{equation}
\begin{figure}[htbp]
    \centering
    \includegraphics[width=1.0\columnwidth]{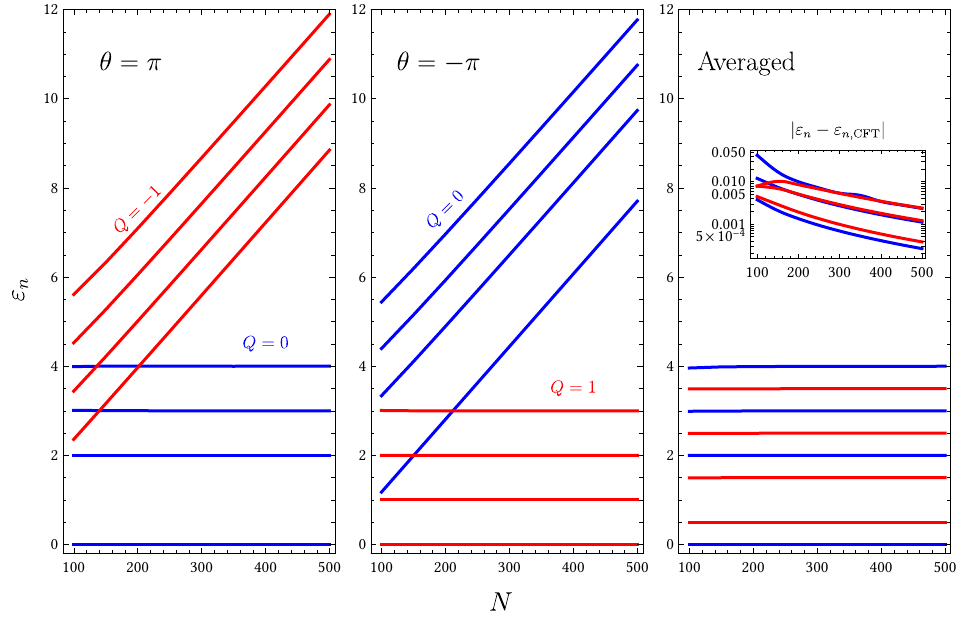}
    \caption{Rescaled energy spectrum of OBC Schwinger Model at the critical point, for a range of lattice sites $N$. The $Q=0$ sector has color blue and $Q=\pm 1$ sector has color red. The plot shows 4 eigenvalues for each $Q$ sector. The definition of vertical axis is $\varepsilon_n = \frac{E_n-E_0}{E_1-E_0} \times \delta$ where $\delta = 2, 1, {\rm~and~} \frac{1}{2}$ for 
    % $L_{\rm in}=0$, 
    % $L_{\rm in}=-1$ 
    $\theta = \pi$, $\theta = -\pi$
    and $\rm Averaged$ (\ref{SpecAv}), respectively, in order to rescale the energy levels to match FIG. \ref{IsingCFT_BC}. The inset in the 3rd subfigure shows the relative error between the numerical energy levels rescaled and the prediction from CFT.
    }
    \label{fig:figCriticalSpectrumOBC}
\end{figure}
with the identification $q = e^{-\beta/2}$ and $E = h$. 
These partition function, when expanded in power series in $q$, provide the energy levels and degeneracy data for FIG. \ref{IsingCFT_BC}. A particular interesting feature in the OBC family is that $F$ has two primaries, $\mathbbm{1}$ and $\epsilon$. Despite its name, $\epsilon$ is parity odd. Thus $(++)$ and $(+-)$ can be obtained by projecting $F$ to corresponding parity sector. In OBC Schwinger model, the $Q=0$ and $Q=-1$ ($Q=+1$) are the two low-lying charge sectors for 
$\theta = \pi$ ($\theta = -\pi$).
The two different charge sectors do not mix with each other, and at the critical point they must flow to separate CFT primaries. Depending on the 
theta angle $\theta = \pm \pi$,
the two sectors get a positive or negative relative $O(1)$ shift, and the higher one is gapped out in the thermodynamic limit, matching $(++)$ or $(+-)$ boundary conditions. With the averaged spectrum (\ref{SpecAv}), the $O(1)$ shift is removed, and the critical spectrum has both primaries, matching the $F$ boundary condition. We check the picture numerically to high precision, and we obtain critical spectrum and degeneracies perfectly matching the CFTs, shown in FIG. \ref{fig:figCriticalSpectrumOBC}.

\section{Numerical results for PBC}

In this section we discuss the numerical result for PBC. 
The procedure of locating the critical point for PBC is the same as in previous sections. We first use a criterion to locate a pseudo critical coupling $m_{*}/e$ at finite $N$ and $x$, extrapolate to the thermodynamic limit $N\rightarrow \infty$ to obtain the true critical coupling $m_{c}/e$ for the finite $x$ lattice model, and finally extrapolate to the continnum limit $x\rightarrow \infty$. 
The criterion we use is similar to $C_2$ defined in (\ref{CritCrPBCLeading}).
As shown in FIG. \ref{IsingCFT_BC}, the ratio between the two lowest excited states in PBC should be $8$: 
\begin{align}
C_{2}^{\rm PBC}: \quad \frac{E_{20}(m_{*}/e, x,N)}{E_{10}(m_{*}/e, x,N)} =  \frac{\Delta_{\epsilon}}{\Delta_{\sigma}} = 8\,.
\label{CritCrPBCLeading}
\end{align}
Nevertheless, in case of PBC we have more information from the CFT side which we can leverage to improve the numerical result. 
The leading finite size effect can be computed through conformal perturbation theory and be subtracted off, resulting in $m_*/e$ measurements that converge at a higher power in $1/N$. 
In Appendix. \ref{sec:CPT} we perform a conformal perturbation theory analysis similar to the analysis in \cite{Lao:2023zis}, and we sketch the main result that we use in the numerical procedure in the following.
\begin{table}[htbp]
\centering

$
\begin{array}{c|c|c|c}
\hline 
X & \Delta_X & \langle X | T\bar T | X \rangle & \langle X | \frac{1}{2}\left(\chi_{0,4}+\bar\chi_{0,4}\right) | X \rangle \\ 
\hline 
\sigma & \frac{1}{8} & \frac{1}{64} & \frac{3}{64 \sqrt{10}} \\ 
\hline 
\epsilon & 1 & 1 & \frac{1}{\sqrt{10}} \\ 
\hline 
\partial\sigma & \frac{9}{8} & \frac{17}{64} & \frac{147}{64 \sqrt{10}} \\ 
\hline 
\partial\epsilon & 2 & 3 & \frac{7}{\sqrt{10}} \\ 
\hline 
T & 2 & 0 & \frac{7}{\sqrt{10}} \\ 
\hline 
\end{array}
$
\caption{\label{tab:OPE-coeff} Table of OPE coefficients involved in (\ref{linearConformalPert}) for some low lying states. 
}
\end{table}
We sketch the key results as the following. We model the critical Hamiltonian as the CFT dilatation operator perturbed by irrelevant local operators. In the leading order perturbation theory, the correction of an energy eigenvalue is the following
\begin{widetext}
\begin{equation}
\label{linearConformalPert}
\frac{N}{\nu} (E_X - E_0) = \Delta_X + \frac{\mu_1}{N^2} \langle X | T\bar T | X \rangle + \frac{\mu_2}{N^2} \langle X | \frac{1}{2}\left(\chi_{0,4}+\bar\chi_{0,4}\right) | X \rangle + O(N^{-2})~. 
\end{equation}
\end{widetext}
The values of the matrix elements $\langle X | \mathcal{O} | X \rangle$ are given by OPE coefficients in the underlying CFT. They are summarized in TABLE \ref{tab:OPE-coeff}. The unknown coefficients $\mu_1$ and $\mu_2$ are model dependent. If we compute the energy of a few more eigenstates, we can find a linear combination where the leading correction cancel out, and we can use it to define an improved criterion: 
\begin{widetext}
    \begin{equation}
        C_{5}: \quad 14 \times \left(\frac{E_{20}(m_{*}/e, x,N)}{E_{10}(m_{*}/e, x,N)} - \frac{\Delta_{\epsilon}}{\Delta_{\sigma}}\right) + 12 \times \left(\frac{E_{30}(m_{*}/e, x,N)}{E_{10}(m_{*}/e, x,N)} - \frac{\Delta_{\partial\sigma}}{\Delta_{\sigma}}\right) - 5 \times \left(\frac{E_{50}(m_{*}/e, x,N)}{E_{10}(m_{*}/e, x,N)} - \frac{\Delta_{\partial\epsilon}}{\Delta_{\sigma}}\right) = 0 \,.
\label{CritCrPBCSubleading}
    \end{equation}
\end{widetext}
The expected convergence of the pseudo critical coupling is
\begin{equation}
    \begin{aligned}
        C_{2}^{\rm PBC}: \quad m_c(x) &= m_*(x) + O(N^{-3}) \\
        C_5: \quad m_c(x) &= m_*(x) + O(N^{-4}) 
    \end{aligned}
\end{equation}
We compute most of the PBC data using $C_5$.
DMRG is slower in PBC and within bond dimension $m\leq 3000$ it can only go up to $N \sim 80$ with reasonable truncation error $(10^{-10})$, which in turn puts a limit on the maximal $x$ which $m_c(x)/e$ can be determined.
\begin{figure}[htpb]
    \centering
    \includegraphics[width=1.0\linewidth]{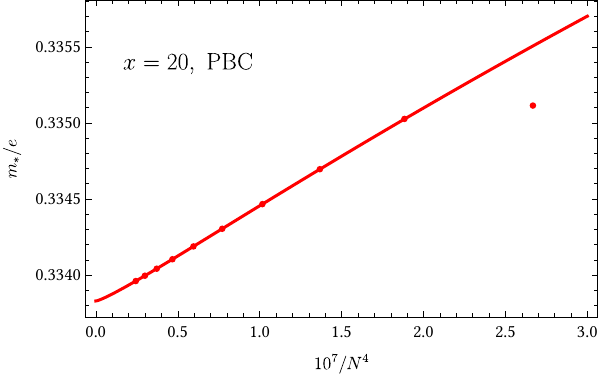}
    \caption{Fit of PBC pseudo critical mass $m_*(x,N)/e$ at $x=20$ determined by the improved PBC criterion $C_5$ (\ref{CritCrPBCSubleading}).}
    \label{fig:pltPBCFitN20}
\end{figure}
For example, the pseudo critical point $m_*(x,N)/e$ data for $N = 20$ is shown in FIG. \ref{fig:pltPBCFitN20}. We use a fit to extrapolate it to the thermodynamic limit
\begin{equation}
\begin{aligned}
    m_*(20,N)/e &= 0.333828\, -\frac{2043.55}{N^4} +\frac{9.71\times10^5}{N^5} \\ 
    &~~-\frac{3.46\times 10^7}{N^6}+\frac{3.54\times 10^8}{N^7}~.
\end{aligned}
\end{equation}
Since $N$ is not large enough, we find that the fit still significantly depends on the high powers in the fit.
Similarly, we computed $m_c(x)/e$ for different $x$ up to $x=24$, and summarize the result in TABLE \ref{tab:PBC_val_table}.
\begin{table}[htbp]
    \centering
    $
        \begin{array}{|c|c|}
        \hline
        x & m_{c}/e, C_5 \\
        \hline
         8 & 0.334238 \\
        \hline
         10 & 0.334099 \\
        \hline
         12 & 0.334007 \\
        \hline
         14 & 0.333943 \\
        \hline
         16 & 0.333895 \\
        \hline
         18 & 0.333858 \\
        \hline
         20 & 0.333828 \\
        \hline
         24 & 0.333784 \\
        \hline
        \end{array}
    $
    \caption{The critical mass value in PBC determined by the improved PBC criterion $C_5$ (\ref{CritCrPBCSubleading}).}
    \label{tab:PBC_val_table}
\end{table}
\begin{figure}[htpb]
    \centering
    \includegraphics[width=1.0\linewidth]{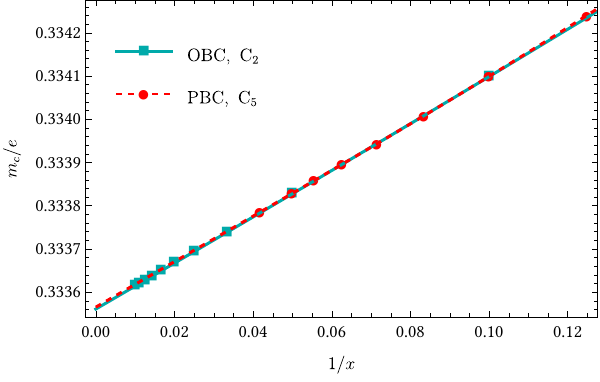}
    \caption{Fit of PBC critical mass $m_c(x)/e$ as a function of $x$. }
    \label{fig:pltPBCFixX}
\end{figure}
We fit the $m_c(x)/e$ to a polynomial in $1/x$. The leading parameters of the fit are similar to the OBC fit. The fit curves shown in FIG. \ref{fig:pltPBCFixX} agree in overlapping regime. The final extrapolation to $x\rightarrow \infty$ agrees with OBC to 5th digit. 
\begin{equation}
    m_c(x)/e = 0.333565 + \frac{0.00527}{x} - \frac{0.00093}{x^2} + \frac{0.0157}{x^3}\,.
\end{equation}
We conclude that PBC result is consistent with the OBC result, though the PBC result may have significant error due to lacking of high-accuracy data for large $x$.

\section*{Acknowledgments}  

We are grateful to Alexey Cherman, Ross Dempsey, Igor R. Klebanov, Vlad Kozii, Shu-Heng Shao, Bernardo Zan, Yijian Zou  for very useful discussions.  We would also like to thank Igor R. Klebanov for his valuable comments on the draft. We also thank Aditya Sinha for collaboration during the early stages of this project.
This work was supported by the Simons Foundation Grant No. 994316. This work used bridges2 cluster at PSC through allocation PHY240237 from the Advanced Cyberinfrastructure Coordination Ecosystem: Services\& Support (ACCESS) program, which is supported by U.S. National Science Foundation grants \#2138259, \#2138286, \#2138307, \#2137603, and \#2138296.

\section*{Appendix}

\subsection{Conformal Perturbation Theory}
\label{sec:CPT}

The lattice model with periodic boundary condition preserves translational symmetry. Compared with the open boundary condition, the spectrum is cleaner and the irrelevant operators involved in the perturbation are more sparse.

We consider the perturbed CFT action 
\begin{equation}
S = S_{\rm CFT} + \sum_i \int d^2 x \mathcal{O}_i(x) \, .
\end{equation}
There are many constraints to the deformation operators
\begin{itemize}
\item No total derivative operators. From the action point of view, deformations of the Lagrangian that are total derivatives are redundant. From the Hamiltonian perspective, a total time derivative is equivalent to a unitary transformation $H \rightarrow e^{-i V} H e^{i V} = H + i[H,V] = H + \partial_\tau V$ expanded to the first order, and thus does not change the spectrum. Derivative is represented by $L_{-1}$ and $\bar{L}_{-1}$. Any global descendant operators with $L_{-1}$ or $\bar{L}_{-1}$ in the front will vanish. 
\item Similarly, deforming by the stress tensor itself is a uniform rescaling of energy and does not change the ratio of energies.
\item No spacial spin component. Such deformation violates translation invariance. There can still be spinning deformation with polarization in the time direction, for example, $(L_{2} + \bar{L}_{2})\mathcal{O}$.
\item For a $\mathbb{Z}_2$-preserving flow, $\sigma$ and its descendants cannot appear in the spectrum. 
\item Time reversal symmetry. Time reverse acts as $(z,\bar{z})\rightarrow (-\bar{z},-z)$, so operators such as $(L_{n} + \bar{L}_{n})\mathcal{O}$ will acquire a sign $(-1)^n$ and odd operators are not allowed in the deformation spectrum.
\end{itemize}
In 2D Ising CFT, all operators are Virasoro descendants of three primary operators $\mathbbm{1}$, $\epsilon$ and $\sigma$. The strategy is to first find the spectrum of the chiral Virasoro descendants of the three primaries and project out total derivatives and null operators. The result is shown in Table \ref{tab:chiral-descendants}. Then we multiply the chiral and anti-chiral Viraosoro generators that survive and further apply translational invariance and time-reversal symmetry. The result is shown in Table \ref{tab:scalar-descendants}.
\begin{table}[htbp]
$
\begin{array}{|c|l|l|l|}
\hline
\text{chiral level} & \mathbbm{1} & \sigma & \epsilon \\
\hline
 2 & T = 2L_{-2} &  &  \\
\hline
 3 & & \chi_{\frac{1}{16},3} &  \\
\hline
 4 & \chi_{0,4} & & \chi_{\frac{1}{2},4} \\
\hline
\end{array}
$
\caption{\label{tab:chiral-descendants}
Left: Chiral quasi-primary operators in Ising CFT. Right: The operators expressed as Virasoro descendants of primary operators.
}
\end{table}
\begin{equation}
\begin{aligned}
\chi_{0,4} &= \frac{2}{7} \sqrt{10} L_{-2}^2-\frac{6}{7} \sqrt{\frac{2}{5}} L_{-4} \\ 
\chi_{\frac{1}{16},3} &= \frac{256}{21} \sqrt{\frac{2}{51}} L_{-1}^3-\frac{2}{7} \sqrt{102} L_{-3} \\ 
\chi_{\frac{1}{2},4} &= \frac{5 L_{-1}^4}{24}-2 L_{-4}
\end{aligned}
\end{equation}

\begin{table}[htbp]
$
\begin{array}{c|l|l|l}
\hline
\text{total level} & \mathbbm{1} & \sigma & \epsilon \\
\hline
 2 &  &  &  \\
\hline
 3 & &  &  \\
\hline
 4 & T\bar T,~ (\chi_{0,4} + \bar\chi_{0,4})/2 & & (\chi_{\frac{1}{2},4} + \bar \chi_{\frac{1}{2},4})/2 \\
\hline
\end{array}
$
\caption{\label{tab:scalar-descendants}
Quasi-primary operators from assembling chiral- and anti-chiral components in Table \ref{tab:chiral-descendants} that have no spacial spin component and is time reversal even.
}
\end{table}

The conformal perturbation Hamiltonian is the following
\begin{equation}
H_{\rm model} = - R \varepsilon_0 + \frac{2\pi}{\nu R} \left( 
L_0 + \bar L_0 - \frac{c}{12}
\right) + \delta H
\end{equation}
where $\nu$ is the speed of light and $\varepsilon_0$ is the vacuum energy. In general $\nu$ and $\varepsilon_0$ can be any positive number depending the details of the UV theory and is not captured by the CFT, and in addition $\nu$ may depend on $N$, so we need to subtract the ground state and take ratios to remove dependence on these quantities. The deformation is given by
\begin{equation}
\delta H = \sum_i \int_0^{2\pi} d\theta \tilde\mu_i \left(\frac{R}{2\pi }\right)^{1-\Delta_i} \mathcal{O}_i(e^{i\theta})
\end{equation}
Since the operators are rotation invariant, we have
\begin{equation}
H_{\rm model} = \frac{2\pi}{R} \left( 
L_0 + \bar L_0 - \frac{c}{12} + \sum_i \frac{\mu_i}{R^{\Delta_i - 2}} \mathcal{O}_i(1)
\right)
\end{equation}
where $\mu_i \equiv (2\pi)^{\Delta_i - 1} \tilde\mu_i$ absorbes the $(2\pi)$ factors.
From Table \ref{tab:scalar-descendants} the leading deformation is at $R^{-2} \sim N^{-2}$ order from two operators at $\Delta = 4$ in the $\mathbbm{1}$ family. Additionaly, at $R^{-3} \sim N^{-3}$ order we see the contribution from a dimension-5 operator in the $\epsilon$ family. The prediction from first-order time-independent perturbation theory gives, for a generic eigenstate $| X \rangle$
\begin{widetext}
\begin{equation}
\tag{\ref{linearConformalPert}}
\frac{N}{\nu} (E_X - E_0) = \Delta_X + \frac{\mu_1}{N^2} \langle X | T\bar T | X \rangle + \frac{\mu_2}{N^2} \langle X | \frac{1}{2}\left(\chi_{0,4}+\bar\chi_{0,4}\right) | X \rangle + O(N^{-3}) 
\end{equation}
\end{widetext}
For some low lying states, their OPE coefficients involved in these orders are shown in Table \ref{tab:OPE-coeff}.

The benefit of the above analysis is that we can write down corrected observables that converges to the critical values faster. For example, the ratio between two energy levels 
\begin{equation}
r_{21} = \frac{E_2 - E_0}{E_1 - E_0} = \frac{\Delta_\epsilon}{\Delta_\sigma} + O(N^{-2}) = 8 + O(N^{-2}) 
\end{equation}
can be used to find critical coupling $g_{c}$ for a given lattice model by finding some $g_{*}$ such that $r_{21} = 8$. The error of this approximation of $g_{c}$ is of order $N^{-3}$, because the dependence of $r_{21}(g,N)$ on $g$ is $\sim(g-g_{c}) N$, assuming that we are sufficiently close to the true, infinite volume critical point such that $g-g_{c} \ll N^{-1}$. From (\ref{linearConformalPert}) and Table \ref{tab:OPE-coeff} we can further construct the quantity
\begin{equation}
\tilde r_{21} \equiv r_{21} + \frac{6}{7}(r_{31} - 9) - \frac{5}{14}(r_{41} - 16) = 8 + O(N^{-3})
\end{equation}
where $r_{31} \equiv \frac{E_3 - E_0}{E_1 - E_0} = \frac{\Delta_{\partial\sigma}}{\Delta_\sigma} + O(N^{-2})$ and $r_{41} \equiv \frac{E_4 - E_0}{E_1 - E_0} = \frac{\Delta_{\partial\epsilon}}{\Delta_\sigma} + O(N^{-2})$ are the ratio involving other energy levels. One can check that the leading finite-volume error cancels out in $\tilde r_{21}$ and the ratio converges faster to the thermodynamic limit.

\subsection{ iTensor implementation for the Schwinger model}
For the efficient use of iTensor  \cite{itensor, itensor-r0.3} for the Schwinger model, we implemented two key features: $U(1)$ charge conservation and explicit MPO construction  \cite{McCulloch_2007, PhysRevA.78.012356, PhysRevB.78.035116} for the Schwinger model Hamiltonian.
The $H_{\textrm{M}}$ and $H_{\textrm{XX}}$  terms in (\ref{HMOBC}) and (\ref{HXXOBC})  can be written as 
\begin{align}
&H_{\textrm{M}} = v^{T}_{\textrm{in}}W^{(0)}_{\textrm{M}} W^{(1)}_{\textrm{M}}   \dots W^{(N-1)}_{\textrm{M}} v_{\textrm{fin}}\,, \\
&H_{\textrm{XX}} = v^{T}_{\textrm{in}}W^{(0)}_{\textrm{XX}} W^{(1)}_{\textrm{XX}}   \dots W^{(N-1)}_{\textrm{XX}} v_{\textrm{fin}}\,,
\end{align}
where the MPO matrices $W_{\textrm{M}}^{(n)}$ and $W_{\textrm{XX}}^{(n)}$ are
\begin{align}
W^{(n)}_{\textrm{M}} = \left(\begin{array}{cc}
    I_{n} & \frac{1}{2}(-1)^{n}Z_{n} \\ 
    0 & I_{n} \\ 
  \end{array}\right)
\end{align}
and
\begin{align}
W^{(n)}_{\textrm{XX}} = \left(\begin{array}{cccc}
    I_{n} & \sigma_{n}^{+} &  \sigma_{n}^{-}& 0 \\ 
    0 & 0 & 0& \sigma_{n}^{-}\\ 
     0 & 0 & 0 &  \sigma_{n}^{+}\\ 
      0 & 0 & 0 & I_{n} \\ 
  \end{array}\right)
\end{align}
with $v_{\textrm{in}}=(1,0,\dots 0)^{T}$ and  $v_{\textrm{fin}}=(0,\dots0, 1)^{T}$ and $\sigma_{n}^{\pm} = (X_{n}\pm i Y_{n})/2$  and $I_{n}$ is $2\times 2$ identity matrix. Finally the non-local Coulomb energy Hamiltonian $H_{\textrm{C}}$ in (\ref{HCOBC}) can be written in the MPO form as 
\begin{align}
&H_{\textrm{C}} = \tilde v^{T}_{\textrm{in}}W^{(0)}_{\textrm{C}} W^{(1)}_{\textrm{C}}   \dots W^{(N-1)}_{\textrm{C}} v_{\textrm{fin}}\,, 
\end{align}
where 
\begin{align}
W^{(n)}_{\textrm{C}} &= \left(\begin{array}{ccc}
    I_{n} & Q_{n} & (N-n-1) Q_{n}^2\\ 
    0 & I_{n} & 2(N-n-1)Q_{n}\\ 
    0 & 0 & I_{n} \\ 
  \end{array}\right)\,, \\ 
\tilde v^{T}_{\textrm{in}} &= \left( \begin{array}{ccc}
    1 & \frac{\theta}{2\pi} &  (N-1) \left(\frac{\theta}{2\pi}\right)^2
\end{array}\right)\,,
\end{align}
where $Q_{n} =\frac{1}{2}(Z_{n}+(-1)^n)$ and 
note that we take $\tilde v_{\textrm{in}}$ to be slightly different from the usual OBC convention in order to include $\theta$ angle.

\subsection{Our  DMRG code implementation for the Schwinger model}
\begin{figure}[h!]
\includegraphics[width=0.35\textwidth]{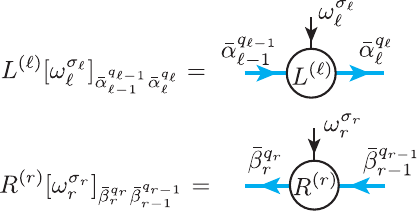}
\caption{Left and Right MPS matrices $L$ and $R$  with their indices for different local $U(1)$ charge sectors. Black lines represent physical bonds, while blue lines denote virtual bonds. The local $U(1)$ charge conservation is expressed as $q_{\ell-1}+\sigma_{\ell}=q_{\ell}$ and $q_{r-1}+\sigma_{r}=q_{r}$. }
\label{DMRG_LandRspin}
\end{figure}

We developed a DMRG code using MATLAB. The details of the implementation, along with the code itself, will be published separately. Similar to the iTensor implementation discussed above, we accelerate the computation by utilizing 
$U(1)$ charge conservation. This is achieved by using MPS matrices with indices corresponding to different local $U(1)$ charge sectors \cite{SCHOLLWOCK201196}, as illustrated in FIG. \ref{DMRG_LandRspin} for the left and right matrices 
$L$ and $R$.
To accelerate the sweeping procedure, we employ the algorithm proposed in \cite{PhysRevLett.77.3633}. Our DMRG code begins with the chain-building phase,  reaching the final number of qubit sites $N$ before proceeding with the sweeping process \cite{PhysRevLett.69.2863, PhysRevB.48.10345}. We observed that the chain-building phase significantly improves the convergence of sweeps compared to the iTensor code, where one typically starts with random MPS matrices, requiring tens of sweeps for convergence. For a single sweep of the Schwinger model with given parameters, our code is typically twice as fast as iTensor. Additionally, the larger the bond dimension, the greater the speed advantage of our code over iTensor. 
For example, computing the ground state energy of the massive Schwinger model in OBC for
$N = 1000$ qubits, achieving machine precision accuracy, takes approximately 
$120$ seconds for the build phase and $100$ seconds for a single sweep on a MacBook Pro M3 laptop.

\bibliographystyle{ieeetr} 
\bibliography{schw_dmrg}

\end{document}